\documentclass[pdflatex,sn-basic]{sn-jnl}

\usepackage{txfonts}
\usepackage{longtable}
\usepackage{subcaption}
\newcommand{\aap}{Astron. Astrophys.} 
\newcommand{\apjs}{Astrophys. J. Suppl. Ser.} 
\jyear{2022}%

\begin{document}

\title[Broad band flux-density monitoring of radio sources with the Onsala twin telescopes]{Broad band flux-density monitoring of radio sources with the Onsala twin telescopes}


\author*[1]{\fnm{Eskil} \sur{Varenius}}\email{eskil.varenius@chalmers.se}

\author[2]{\fnm{Francesco} \sur{Maio}}

\author[1]{\fnm{Karine} \sur{Le Bail}}

\author[1]{\fnm{R\"udiger} \sur{Haas}}

\affil*[1]{\orgdiv{Department of Space, Earth and Environment}, \orgname{Chalmers University of Technology}, \orgaddress{\street{Onsala Space Observatory}, \city{Onsala}, \postcode{43992}, \country{Sweden}}}

\affil[2]{\orgdiv{Enrico Fermi Department of Physics}, \orgname{University of Pisa}, \orgaddress{\street{Largo Bruno Pontecorvo 3}, \city{Pisa}, \postcode{56127},  \country{Italy}}}


\abstract{\textbf{Context and aims:} The Onsala twin telescopes (OTT) are two 13~m telescopes located at the Onsala Space Observatory in Sweden. With dual linear polarized broad-band (3-14~GHz) receivers, they are part of the next generation Very Long Baseline Interferometry (VLBI) Global Observing System (VGOS) for geodesy and astrometry. In addition to purely geodetic data products, VGOS will regularly produce full-polarisation images of hundreds of radio sources. These rich monitoring data will be valuable for both astronomy and geodesy. In this pilot study we aim to monitor 10 bright radio sources to search for flares or similar activity, and to verify the instrument calibration on long (months) and short (hours) time scales. 

\textbf{Method:} We observed and analysed 91 short (\textless30~min) sessions spanning 7 months. We monitored seven potentially variable radio sources (0059+581, 0552+398, 1144+402, 1156+295, 1617+229, 3C418, OJ287) and three reference calibrators (3C147, 3C286, 3C295). We used the Common Astronomy Software Applications (CASA) package to fringe-fit, bandpass-correct and scale the data to obtain flux densities in the four standard VGOS bands: 3.0-3.5 GHz (band~1),  5.2-5.7 GHz (band~2), 6.3-6.8 GHz (band~3), and 10.2-10.7 GHz (band~4).
\textbf{Results:} We obtain simultaneous multi-frequency light curves for ten radio sources. A bright multi-frequency flare is observed in the radio source 0059+581. OJ287 and 1156+295 show significant long-term variability.
\textbf{Conclusions:} After correcting for instrumental biases, we determine the empirical flux density uncertainty as $\sim$5~\%. Future refined analysis methods will allow regular monitoring of more and fainter sources.}

\keywords{telescopes: Onsala twin telescopes --  techniques: interferometric -- galaxies: individual: OJ287, 0059+581, 0552+398, 1144+402, 3C418, 1156+295, 1617+229}



\maketitle

\section{Introduction}
VGOS is an international collaboration to provide highly accurate terrestrial and celestial reference frames, as well as highly accurate Earth orientation parameters (EOP), i.e. polar motion, universal time and celestial pole offsets. Several of the EOPs can only be determined with geodetic VLBI and are essential for the operation of all kind of satellite systems, including the widely used Global Navigation Satellite Systems such as GPS and Galileo. Geodetic VLBI, and thus VGOS, is therefore crucial for all kinds of positioning and navigation applications. For further details about geodetic VLBI in general and VGOS in particular, we refer the reader to \cite{Sovers_et_al_1998, Petrachenko_et_al_2012,Nothagel_et_al_2017} and references therein.

VGOS aims to significantly improve the accuracy of geodetic data products, compared to traditional legacy (S/X-band) observations. This means new technical challenges and requirements, for example the need for continuous monitoring of telescope system temperatures, often referred to as Tsys, during observations. Continuous Tsys monitoring is required to provide up-to-date flux densities for optimal scheduling of VLBI observations \citep{LeBail_et_al_2021}, and to account for source structure effects in geodetic analysis \citep{Shabala_et_al_2014, Xu_et_al_2019}.

The Onsala twin telescopes, ONSA13NE (Oe) and ONSA13SW (Ow), separated by 75 m, are currently the only fully operational VGOS twin telescopes in the world \citep{Haas_et_al_2019}. They offer a unique setup to verify and explore the new broad-band VGOS capabilities in terms of hardware and software. In particular, the OTT are since spring 2021 able to monitor Tsys during observations \citep{LeBail_et_al_2021}. This capability does, however, need to be verified by commissioning observations. While single-dish observations can offer some guidance, proper verification of the system performance require interferometric observations, as would be the case in full VGOS observing. This can be achieved by using the OTT as a single-baseline interferometer, where the output data streams are correlated to form complex visibilities which can then be appropriately calibrated to obtain flux densities of celestial sources. By comparing the results with known reference sources, the stability and accuracy of the telescope observing systems can be assessed. If done successfully, we also obtain (the first) simultaneously measured light curves of sources in the four VGOS bands.

In this paper, we report on the results of commissioning observations carried out with the OTT during 7 months in 2021. In Sect. \ref{sec:obs} we describe the observations and the methods used to analyse the data. The results are presented in Sect.~\ref{sec:results} and are discussed in Sect.~\ref{sec:discussion}. Finally, we present our conclusions in Sect.~\ref{sec:conclusions}.

\section{Observations and data analysis}
\label{sec:obs}
We scheduled 102 short ($<30$~min) experiments from May 2021 to December 2021 including between two and ten sources in each experiment. Two series of experiments were scheduled: the Flux Monitoring (FM) sessions spanning 7 months, and the Intra-Day Variability Monitoring (IDVM) sessions spanning 3 days. A list of all scheduled observations is presented in Table~\ref{tab:observations}. Due to limited observing time and/or correlator time, not all sources were included in all observations. For cross-checking purposes, at least one absolute flux-density reference source was included in all experiments starting with fm2110. Eleven of the scheduled experiments failed in different ways, as noted with comments in Table~\ref{tab:observations}, and were excluded from our analysis. We note that initially the campaign was planned for the period May to August, but a few additional observations were made in December to verify the system performance after maintenance work (see Sects.~\ref{sec:gcomo} and \ref{sec:noisediode}).

The OTT are equipped with DBBC3 backends \citep{2018evn..confE.140T}, sampling the analog waveform and producing VLBI Data Interchange Format (VDIF) packets which we record on disk using jive5ab\footnote{https://github.com/jive-vlbi/jive5ab}. We use the standard VGOS observing setup recording 8 Gbps per telescope. Each telescope recorded 32 intermediate frequencies of 32 MHz width (see Table~\ref{tab:freqs}), both polarizations using 2-bit sampling.

\setlength{\tabcolsep}{0.3em} 
\begin{longtable}{l l | c c c c c c c c c c | l}
\caption{The 102 experiments scheduled in this work. The 11 failed experiments were excluded from analysis.\label{tab:observations}}\\
\hline\hline
Exp. code  &  Start [MJD] & \rotatebox{90}{0059+581} & \rotatebox{90}{0552+398} & \rotatebox{90}{1144+402} & \rotatebox{90}{1156+295} & \rotatebox{90}{1617+229} & \rotatebox{90}{3C418} & \rotatebox{90}{OJ287} & \rotatebox{90}{3C147} & \rotatebox{90}{3C286} & \rotatebox{90}{3C295} & Note if excluded\\
\hline
\endfirsthead
\caption{continued.}\\
\hline\hline
Exp. code  &  Start [MJD] & \rotatebox{90}{0059+581} & \rotatebox{90}{0552+398} & \rotatebox{90}{1144+402} & \rotatebox{90}{1156+295} & \rotatebox{90}{1617+229} & \rotatebox{90}{3C418} & \rotatebox{90}{OJ287} & \rotatebox{90}{3C147} & \rotatebox{90}{3C286} & \rotatebox{90}{3C295} & Note if excluded\\
\hline
\endhead
\endfoot
fm2101  &  59344.5212   & X  & X  & X  & X  &    & X  & X  &    &    &   & No Tsys data\\
fm2102  &  59345.5212   & X  & X  & X  & X  &    & X  & X  &    &    &   & \\
fm2103  &  59346.5532   & X  & X  & X  & X  &    & X  & X  &    &    &   & \\
fm2104  &  59353.5532   & X  & X  & X  & X  &    & X  & X  &    &    &   & \\
fm2105  &  59354.5532   & X  & X  & X  & X  &    & X  & X  &    &    &   & \\
fm2106  &  59355.5532   & X  & X  & X  & X  &    & X  & X  &    &    &   & \\
fm2107  &  59358.5532   & X  & X  & X  & X  &    & X  & X  &    &    &   & \\
fm2108  &  59359.5532   & X  & X  & X  & X  &    & X  & X  &    &    &   & \\
fm2109  &  59360.5532   & X  & X  & X  & X  &    & X  & X  &    &    &   & \\
fm2110  &  59361.5579   & X  & X  & X  & X  &    & X  & X  & X  & X  &   & \\
fm2111  &  59365.5580   & X  & X  & X  & X  &    & X  & X  & X  & X  &   & \\
fm2112  &  59366.5580   & X  & X  & X  & X  &    & X  & X  & X  & X  &   & \\
fm2113  &  59367.5580   & X  & X  & X  & X  &    & X  & X  & X  & X  &   & \\
fm2114  &  59368.5580   & X  & X  & X  & X  &    & X  & X  & X  & X  &   & Band 4 error\\
fm2115  &  59369.5580   & X  & X  & X  & X  &    & X  & X  & X  & X  &   & \\
fm2116  &  59372.5482   & X  & X  & X  & X  &    & X  & X  & X  & X  &   & \\
fm2117  &  59373.5483   & X  & X  & X  & X  &    & X  & X  & X  & X  &   & \\
fm2118  &  59374.5483   & X  & X  & X  & X  &    & X  & X  & X  & X  &   & Ow ACU error\\
fm2119  &  59376.5483   & X  & X  & X  & X  &    & X  & X  & X  & X  &   & \\
fm2120  &  59379.6757   & X  & X  & X  & X  & X  & X  & X  & X  & X  &   & \\
fm2121  &  59380.3405   & X  & X  & X  &    &    & X  & X  & X  &    & X & No Tsys data\\
fm2122  &  59381.4699   & X  & X  & X  & X  &    & X  & X  & X  & X  & X & \\
fm2123  &  59382.6793   & X  & X  & X  & X  & X  & X  & X  & X  & X  & X & \\
fm2124  &  59383.5520   & X  & X  & X  & X  &    & X  & X  & X  & X  & X & \\
fm2125  &  59385.5103   & X  & X  & X  & X  &    & X  & X  & X  & X  & X & \\
fm2126  &  59386.6770   & X  & X  & X  & X  &    & X  & X  & X  & X  & X & \\
fm2127  &  59387.6376   & X  & X  & X  & X  & X  & X  & X  & X  & X  & X & \\
fm2128  &  59388.6793   & X  & X  & X  & X  & X  & X  & X  & X  & X  & X & \\
fm2129  &  59389.5961   & X  & X  & X  & X  & X  & X  & X  & X  & X  & X & \\
fm2130  &  59393.6369   & X  & X  & X  & X  & X  & X  & X  & X  & X  & X & \\
fm2131  &  59394.5538   & X  & X  & X  & X  & X  & X  & X  & X  & X  & X & \\
fm2132  &  59395.5538   & X  & X  & X  & X  & X  & X  & X  & X  & X  & X & \\
fm2133  &  59397.5538   & X  & X  & X  & X  & X  & X  & X  & X  & X  & X & \\
idvm01  &  59397.6340   & X  &    &    & X  &    &    &    &    & X  & X & \\
idvm02  &  59397.7174   & X  &    &    & X  &    &    &    &    & X  & X & \\
idvm03  &  59397.8007   & X  &    &    & X  &    &    &    &    & X  & X & \\
idvm04  &  59397.8840   & X  &    &    & X  &    &    &    &    & X  & X & \\
idvm05  &  59397.9673   & X  &    &    & X  &    &    &    &    & X  & X & \\
idvm06  &  59398.0463   & X  &    &    &    &    &    &    &    & X  & X & \\
idvm07  &  59398.1273   & X  &    &    &    &    &    &    &    &    & X & \\
idvm08  &  59398.2107   & X  &    &    &    &    &    &    &    &    & X & \\
idvm09  &  59398.2942   & X  &    &    &    &    &    &    &    &    & X & \\
idvm10  &  59398.3817   & X  &    &    & X  &    &    &    &    &    & X & \\
idvm11  &  59398.4673   & X  &    &    & X  &    &    &    &    & X  & X & \\
fm2134  &  59398.5538   & X  & X  & X  & X  & X  & X  & X  & X  & X  & X & \\
idvm12  &  59398.6340   & X  &    &    & X  &    &    &    &    & X  & X & \\
idvm13  &  59398.7174   & X  &    &    & X  &    &    &    &    & X  & X & \\
idvm14  &  59398.8007   & X  &    &    & X  &    &    &    &    & X  & X & \\
idvm15  &  59398.8840   & X  &    &    & X  &    &    &    &    & X  & X & \\
idvm16  &  59398.9673   & X  &    &    & X  &    &    &    &    & X  & X & \\
idvm17  &  59399.0463   & X  &    &    &    &    &    &    &    & X  & X & \\
idvm18  &  59399.1273   & X  &    &    &    &    &    &    &    &    & X & \\
idvm19  &  59399.2107   & X  &    &    &    &    &    &    &    &    & X & \\
idvm20  &  59399.2942   & X  &    &    &    &    &    &    &    &    & X & \\
idvm21  &  59399.3795   & X  &    &    & X  &    &    &    &    &    & X & \\
idvm22  &  59399.4673   & X  &    &    & X  &    &    &    &    & X  & X & \\
fm2135  &  59399.5538   & X  & X  & X  & X  & X  & X  & X  & X  & X  & X & \\
idvm23  &  59399.6340   & X  &    &    & X  &    &    &    &    & X  & X & \\
idvm24  &  59399.7174   & X  &    &    & X  &    &    &    &    & X  & X & \\
idvm25  &  59399.8007   & X  &    &    & X  &    &    &    &    & X  & X & \\
idvm26  &  59399.8840   & X  &    &    & X  &    &    &    &    & X  & X & \\
idvm27  &  59399.9673   & X  &    &    & X  &    &    &    &    & X  & X & \\
idvm28  &  59400.0463   & X  &    &    &    &    &    &    &    & X  & X & \\
idvm29  &  59400.1273   & X  &    &    &    &    &    &    &    &    & X & \\
idvm30  &  59400.2107   & X  &    &    &    &    &    &    &    &    & X & \\
idvm31  &  59400.2942   & X  &    &    &    &    &    &    &    &    & X & \\
idvm32  &  59400.3795   & X  &    &    & X  &    &    &    &    &    & X & \\
idvm33  &  59400.4673   & X  &    &    & X  &    &    &    &    & X  & X & \\
fm2136  &  59400.5538   & X  & X  & X  & X  & X  & X  & X  & X  & X  & X & \\
fm2137  &  59402.6161   & X  & X  & X  & X  & X  & X  & X  & X  & X  & X & \\
fm2138  &  59403.5953   & X  & X  & X  & X  & X  & X  & X  & X  & X  & X & \\
fm2139  &  59407.5543   & X  & X  & X  & X  & X  & X  & X  & X  & X  & X & Oe ACU error\\
fm2140  &  59408.5543   & X  & X  & X  & X  & X  & X  & X  & X  & X  & X & \\
fm2141  &  59409.5546   & X  & X  & X  & X  & X  & X  & X  & X  & X  & X & \\
fm2142  &  59410.5546   & X  & X  & X  & X  & X  & X  & X  & X  & X  & X & \\
fm2143  &  59411.5546   & X  & X  & X  & X  & X  & X  & X  & X  & X  & X & \\
fm2144  &  59414.5546   & X  & X  & X  & X  & X  & X  & X  & X  & X  & X & \\
fm2145  &  59415.5755   & X  & X  & X  & X  & X  & X  & X  & X  & X  & X & Low amp.\\
fm2146  &  59416.5546   & X  & X  & X  & X  & X  & X  & X  & X  & X  & X & \\
fm2147  &  59417.5546   & X  & X  & X  & X  & X  & X  & X  & X  & X  & X & \\
fm2148  &  59421.5546   & X  & X  & X  & X  & X  & X  & X  & X  & X  & X & \\
fm2149  &  59422.5546   & X  & X  & X  & X  & X  & X  & X  & X  & X  & X & Low amp.\\
fm2150  &  59423.5546   & X  & X  & X  & X  & X  & X  & X  & X  & X  & X & Low amp.\\
fm2151  &  59424.5538   & X  & X  & X  & X  & X  & X  &    & X  & X  & X & \\
fm2152  &  59428.5538   & X  & X  & X  & X  & X  & X  &    & X  & X  & X & \\
fm2153  &  59429.5538   & X  & X  & X  & X  & X  & X  &    & X  & X  & X & \\
fm2154  &  59430.5538   & X  & X  & X  & X  & X  & X  &    & X  & X  & X & \\
fm2155  &  59443.5538   & X  & X  & X  & X  & X  & X  &    & X  & X  & X & \\
fm2156  &  59445.5548   & X  & X  & X  & X  & X  & X  & X  & X  & X  & X & \\
fm2157  &  59449.5549   & X  & X  & X  & X  & X  & X  & X  & X  & X  & X & \\
fm2158  &  59457.3882   & X  & X  & X  & X  & X  & X  & X  & X  & X  & X & \\
fm2159  &  59457.4299   & X  & X  & X  & X  & X  & X  & X  & X  & X  & X & \\
fm2160  &  59480.5132   & X  & X  & X  & X  & X  & X  & X  & X  & X  & X & Low amp.\\
fm2161  &  59485.5549   & X  & X  & X  & X  & X  & X  & X  & X  & X  & X & Recorder error\\
fm2162  &  59486.5549   & X  & X  & X  & X  & X  & X  & X  & X  & X  & X & Recorder error\\
fm2163  &  59548.3257   & X  & X  & X  & X  & X  & X  & X  & X  & X  & X & \\
fm2164  &  59549.2841   & X  & X  & X  & X  & X  & X  & X  & X  & X  & X & \\
fm2165  &  59550.2841   & X  & X  & X  & X  & X  & X  & X  & X  & X  & X & \\
fm2166  &  59551.3257   & X  & X  & X  & X  & X  & X  & X  & X  & X  & X & \\
fm2167  &  59554.2841   & X  & X  & X  & X  & X  & X  & X  & X  & X  & X & \\
fm2168  &  59555.3049   & X  & X  & X  & X  & X  & X  & X  & X  & X  & X & \\
fm2169  &  59557.3257   & X  & X  & X  & X  & X  & X  & X  & X  & X  & X & \\
\end{longtable}

\begin{table}[htbp]
    \caption{The frequency setup used in standard VGOS observations. Each spectral window is 32 MHz wide and was observed in both horizontal and vertical polarization. Horizontal lines define the four VGOS bands (1) 3.0-3.5 GHz, (2) 5.2-5.7 GHz, (3) 6.3-6.8 GHz, and (4) 10.2-10.7 GHz. The spectral window (SPW) IDs and order is given as reported by CASA, which also corresponds to the SPW ID in the ASCII files with flux density measurements available in the supplementary material of this article.}
    \label{tab:freqs}
    \centering
    \begin{tabular}{c|c}
  SPW & Center freq. [MHz]  \\
   \hline
    \hline
  0     &  3464.35    \\
  1     &  3432.35    \\
  2     &  3368.35    \\
  3     &  3304.35    \\
  4     &  3208.35    \\
  5     &  3080.35    \\
  6     &  3048.35    \\
  7     &  3016.35    \\
  \hline
  8     &  5704.35    \\
  9     &  5672.35    \\
  10    &  5608.35    \\
  11    &  5544.35    \\
  12    &  5448.35    \\
  13    &  5320.35    \\
  14    &  5288.35    \\
  15    &  5256.35    \\
   \hline
  16    &  6824.35    \\
  17    &  6792.35    \\
  18    &  6728.35    \\
  19    &  6664.35    \\
  20    &  6568.35    \\
  21    &  6440.35    \\
  22    &  6408.35    \\
  23    &  6376.35    \\
   \hline
  24    & 10664.35    \\
  25    & 10632.35    \\
  26    & 10568.35    \\
  27    & 10504.35    \\
  28    & 10408.35    \\
  29    & 10280.35    \\
  30    & 10248.35    \\
  31    & 10216.35    \\
    \end{tabular}
\end{table}

\subsection{Sample selection}
We selected seven sources based on a their properties in existing literature. Starting from the VGOS standard source catalog used in regular observations, we used the Bordeaux VLBI Image  Database (BVID)\footnote{http://bvid.astrophy.u-bordeaux.fr/database.html} to identify sources with different brightness and variability, and the NRAO/VLA Sky Survey (NVSS)\footnote{https://www.cv.nrao.edu/nvss/postage.shtml} to identify sources with significant structure and/or other nearby companions. A summary of the properties for the selected sources are given in Table~\ref{tab:sources}. We note that while the OTT baseline length of 75 m in theory can achieve a maximum synthesised resolution of about 3 arc minutes at 5 GHz, the actual projected baseline length may often be significantly shorter. The corresponding maximum field-of view for the 13 m single dish is about 20 arc minutes. \emph{Structure} in Table~\ref{tab:sources} hence mean that we find significant structure on scales of 3 arc minutes or larger in the respective NVSS field image, and \emph{Isolated} means that we find other bright sources within a 10 arc minute radius. We also note that while NVSS is a good match for the OTT angular scales, it is a significantly lower frequency (1.4 GHz) than the VGOS bands. The structure of our selected sources and nearby sources may hence differ at the VGOS bands. Similarly, the BVID flux densities include 8.4 GHz (X-band) measurements which does not overlap directly with the VGOS observing setup. Nevertheless, we selected the seven sources based on these BVID and NVSS data as no corresponding VGOS-catalog exists. 

\begin{table}[htbp]
    \caption{Summary of properties for the selected radio sources. }
    \label{tab:sources}
    \centering
    \begin{tabular}{l|l l l l}
         Source & Brightness  & Structure  & Isolated & Variable  \\
         \hline
         \hline
0552+398 & Bright      & Point      & Yes      & No \\
1144+402 & Medium      & Point      & Yes      & Yes \\
3C418    & Bright      & Extended   & No       & No \\
OJ287    & Bright      & Point      & No       & Yes \\
0059+581 & Bright      & Point      & No       & Yes \\
1156+295 & Medium      & Point      & No       & Yes \\
1617+229 & Faint       & Point      & Yes      & No\\ 
    \end{tabular}
\end{table}

\subsection{Obtaining flux densities from raw telescope data}
We used DiFX 2.6 \citep{Deller2007, Deller2011} to correlate the VDIF data and obtain complex visibilities, which were exported to FITSIDI-format using the DiFX tool \emph{difx2fits}. Since the antennas share the same clock, and normally use a 5 MHz spaced pulse-calibration system for geodetic VLBI corrections, the pulse-cal tones correlate and produce strong interference in the visibility spectra. To allow sharp removal of these tones, and to account for any unknown significant instrumental effects, we correlated with high (0.1 MHz=320 channels per SPW) spectral resolution, and 1 second time resolution. Astronomical VLBI telescopes often distribute Tsys measurements in antab format\footnote{http://www.aips.nrao.edu/cgi-bin/ZXHLP2.PL?ANTAB}. To make use of existing software, we hence stored our Tsys measurements (and gain curve models) as antab files. The antab information (Tsys+gain) was then appended to the FITSIDI using the CASA VLBI Tools developed by JIVE\footnote{https://github.com/jive-vlbi/casa-vlbi}.

We developed a calibration script in python to process all our data using CASA tools \citep{CASA}. First, the data were imported to Measurement Set (MS) format using the CASA task \emph{importfitsidi}. We used \emph{accor} to normalise the visibility amplitudes using auto-correlations, \emph{fringefit} to determine delay, rate and phase corrections (with 30 sec solution interval, one solution per SPW per polarisation), and \emph{bandpass} to adjust for instrumental bandpass effects (using 3C295 or, if missing, instead OJ287). In addition to removing the predictable pulse-cal tones, the task \emph{rflag} was used to mitigate other radio frequency interference (RFI), in particular in the lowest band~1. For the interested reader, the full python script used for the CASA processing (the same for all experiments) is available as online material attached to this paper. 

\begin{figure}[htb]
     \includegraphics[width=0.48\textwidth]{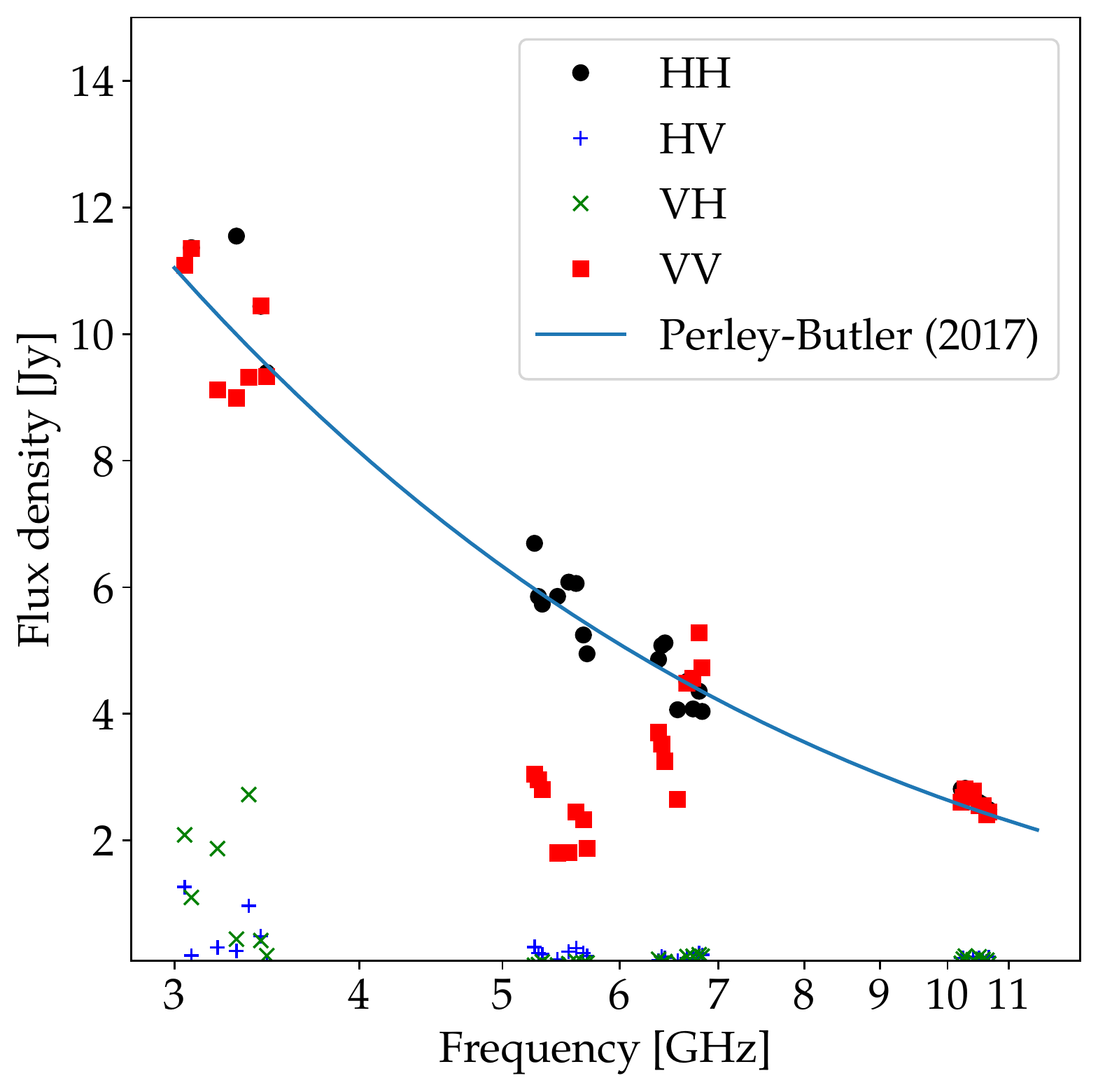}
     \caption{The measured full-polarisation spectrum, per spectral window, as obtained in experiment fm2142 for the absolute flux density reference source 3C295. The solid line shows the model from \cite{2017ApJS..230....7P}. Some of the points in band~1 (3.0-3.5 GHz) are outside the scale of this plot, likely due to residual RFI. The non-zero cross-polarisation amplitudes (HV and VH) in band~1 is also likely (at least partially) due to RFI. A significant offset of the VV-flux densities in band~2 (5.2-5.7 GHz) is clear, and was significantly improved by repairing a faulty down-conversion board (see Sect. \ref{sec:gcomo}).}
     \label{fig:fm_per_spw}
\end{figure}

After processing, the data were averaged to obtain, for each source, one flux density value per SPW and polarisation. All values obtained for all experiments are available as online material attached to this paper. An example is shown in Fig.~\ref{fig:fm_per_spw}. While there is generally good agreement between the measured values and values from the literature for the three flux density calibrators, a few particular aspects needed further attention.

\subsection{Empirical scaling factors}
\label{sec:gcomo}
From first analysis it was clear that band~2 VV-polarisation amplitudes were consistently lower than expected, see Fig.~\ref{fig:fm_per_spw}. This was eventually determined to be due to a faulty down-conversion board in the Oe DBBC3. We therefore determined an empirical correction factor for band~2 by comparing our data with the \citep{2017ApJS..230....7P} models for 3C147, 3C286 and 3C295. After repairing the board, in September 2021, the VV-level (from fm2160) improved significantly. The scaling factors applied to our band~2-data were 1.4 before the repair, and 1.1 after the repair.

To account for other apparent systematics we also scaled, again from comparing with the reference source models, band~3 by a factor 1.05 before fm2160 and band~4 by a factor 0.95 after fm2160.

These empirical scaling factors have been included in the figures presented in this paper, but not in the machine readable tables of flux density values available electronically.

\subsection{Noise diode power supply failures}
\label{sec:noisediode}
In September 2021 the noise diode signals of both telescopes failed within a few weeks of each other. First gradually, with unstable levels and non-linear on/off behaviour, and then completely with no signal received. The problem was faulty power supplies used for sending the on/off signal from the control building to the telescopes, and nothing faulty with the actual noise sources themselves. Some experiments which did not process as expected could perhaps have been affected by the first gradually failing stage of these power supplies, these are noted as \emph{Low. amp} in Table~\ref{tab:observations} and excluded from analysis. After replacing the power supplies with new units, the noise reference level returned to its previously stable value. The last experiments, from fm2163, all behaved as expected.

\subsection{Removing outliers and forming Stokes I}
Despite using \emph{rflag} in our CASA processing to mitigate RFI, there are still (broader) RFI signals affecting some SPWs in our data. This is most severe in the lowest band~1, but sometimes also affect higher frequencies. Either directly in the data, or indirectly via the Tsys measurements. We also suspect there may be minor residual instrumental biases for some SPWs, which cause abnormally high or low values. To mitigate the impact of these outliers, we employed two filter conditions when averaging the HH and VV SPW flux densities together to provide one Stokes I value per VGOS band. Firstly, we excluded any SPWs where the cross-polarisation correlation amplitudes (HV or VH) were higher than 15~\% of either HH or VV values. The exception was for 3C286, which is known to be linearly polarised at the 10~\% level, where we empirically (with good results) put the cutoff at 30~\% of HV or VH. Secondly, we calculated the median of the SPWs in each band and excluded values deviating more than a factor of 2 from the median value. This procedure turned out to provide robust results, at the expense of losing some data - in particular in band~1. This was most severe for the weakest source 1617+229 where no usable data were recovered for band~1. However, we decided this loss is acceptable.

\subsection{Estimating uncertainties}
Based on the empirical scatter in our time series for the reference sources, see Fig.~\ref{fig:fm_fluxcal}, we adopt an empirical uncertainty of 5~\% for all bands. With this, we obtain the reduced $\chi^2$ values for our reference calibrator sources given in Table~\ref{tab:chi2}. For most bands, except 3C286 band~1 and 3C295 band~4 (which could potentially be explained by excess emission from other field sources), 5~\% uncertainty appears too pessimistic. However, the antenna gain curves are only known with 5-10~\% uncertainties, and hence residual direction-dependent gain errors could affect our data. Detailed modelling to account for these effects is beyond the scope of this work. For simplicity, we therefore adopt a  5~\% uncertainty for all bands based on our reduced $\chi^2$ analysis.

\begin{table}[htbp]
    \caption{Reduced $\chi^2$ values for the three reference sources, using data from the FM-sessions shown in Fig.~\ref{fig:fm_fluxcal} and the models of \cite{2017ApJS..230....7P}.
    \label{tab:chi2}}
    \centering
    \begin{tabular}{c|c|c|c }
  Band & 3C147 & 3C286 & 3C295  \\
   \hline
    \hline
  1     &  0.89 & 1.16 & 0.69\\
  2     &  0.44 & 0.46 & 0.18\\
  3     &  0.25 & 0.48 & 0.17\\
  4     &  0.21 & 0.33 & 0.92\\
    \end{tabular}
\end{table}

\section{Results}
\label{sec:results}
\begin{figure*}[htbp]
     \centering
     \begin{subfigure}[bt]{0.32\textwidth}
         \centering
         \includegraphics[width=\textwidth]{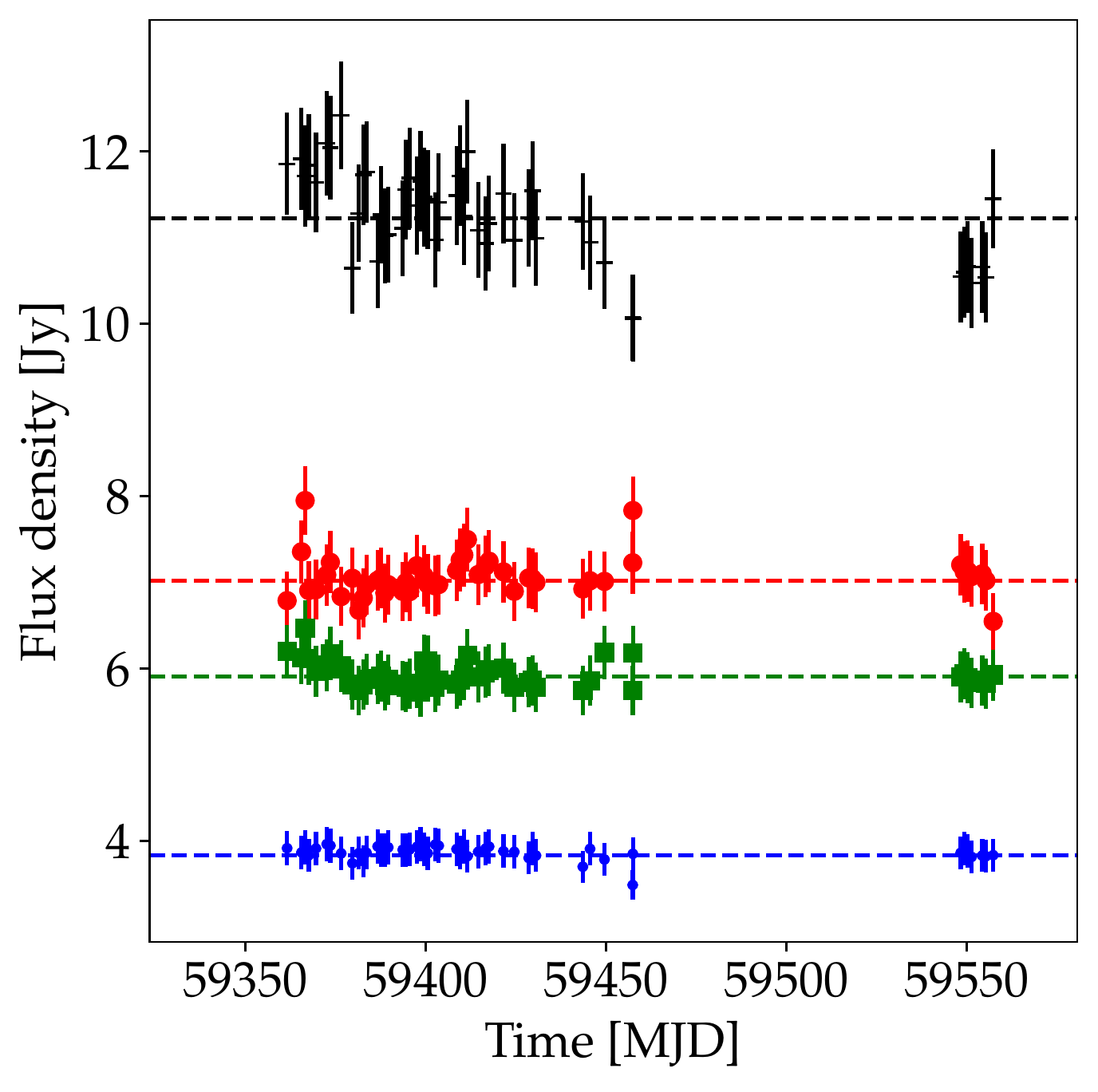}
         \caption{3C147}
         \label{fig:fm_3C147}
     \end{subfigure}
     \hfill
     \begin{subfigure}[bt]{0.32\textwidth}
         \centering
         \includegraphics[width=\textwidth]{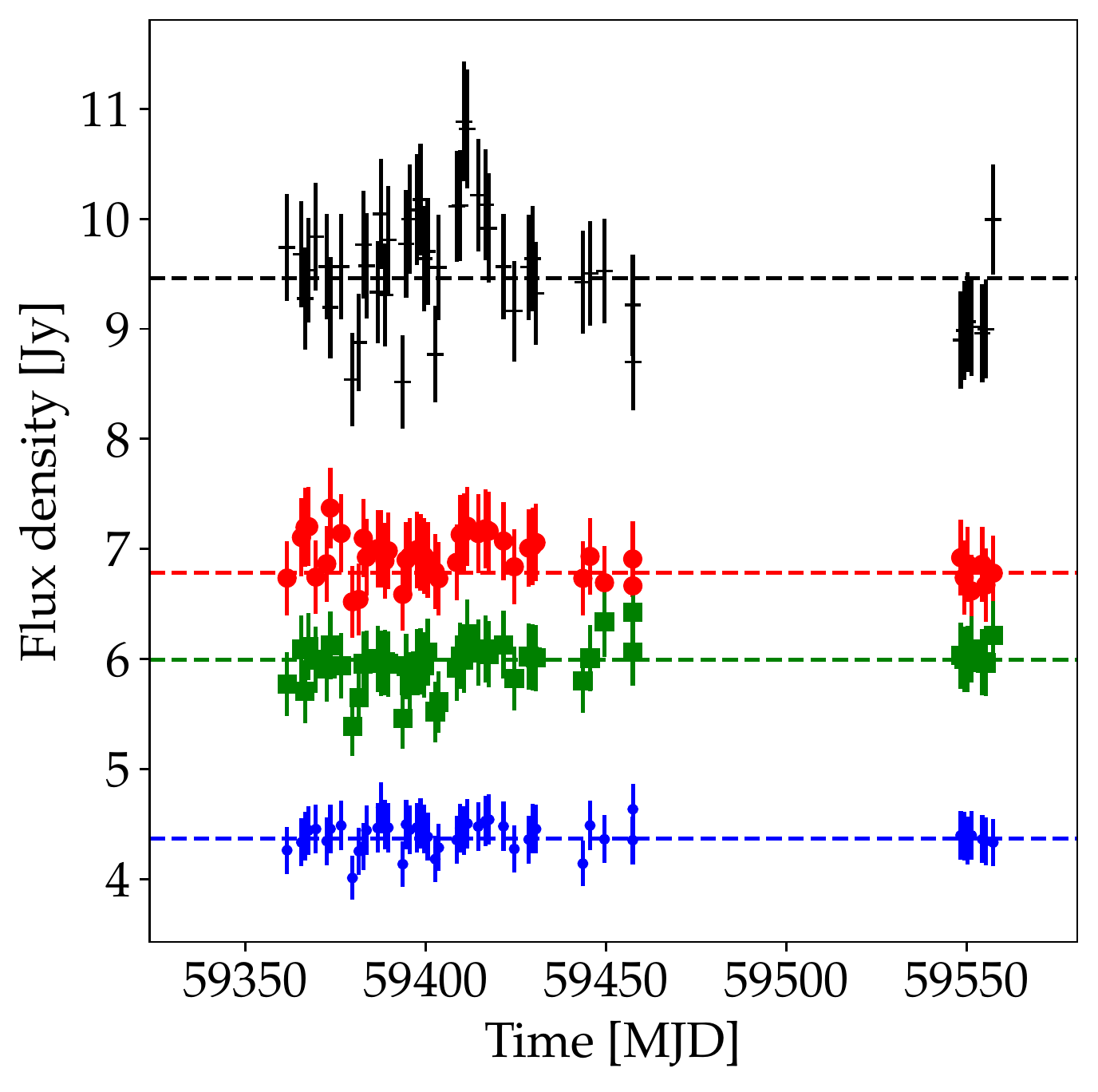}
         \caption{3C286}
         \label{fig:fm_3C286}
     \end{subfigure}
     \hfill
     \begin{subfigure}[bt]{0.32\textwidth}
         \centering
         \includegraphics[width=\textwidth]{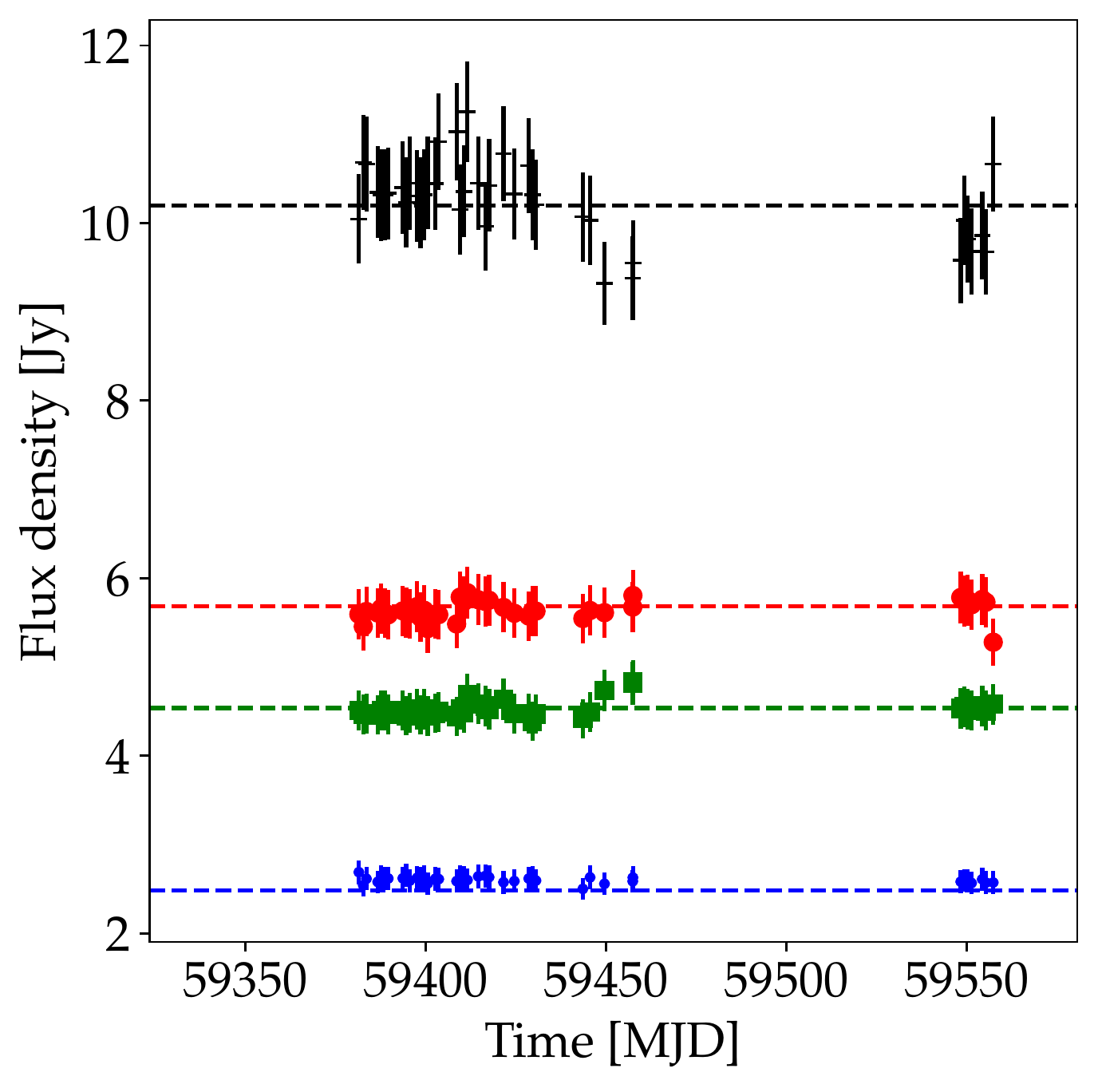}
         \caption{3C295}
         \label{fig:fm_3C295}
     \end{subfigure}
        \caption{Light curves for the three flux density calibrator sources observed in the FM-sessions. The dashed line represent the model from \cite{2017ApJS..230....7P}. Flux densities in the four VGOS bands 1, 2, 3 and 4 are shown with black crosses, red circles, green squares and blue dots respectively. Note that the vertical scale is different for the different sources.}
        \label{fig:fm_fluxcal}
\end{figure*}

\begin{figure*}[htb]
         \includegraphics[width=\textwidth]{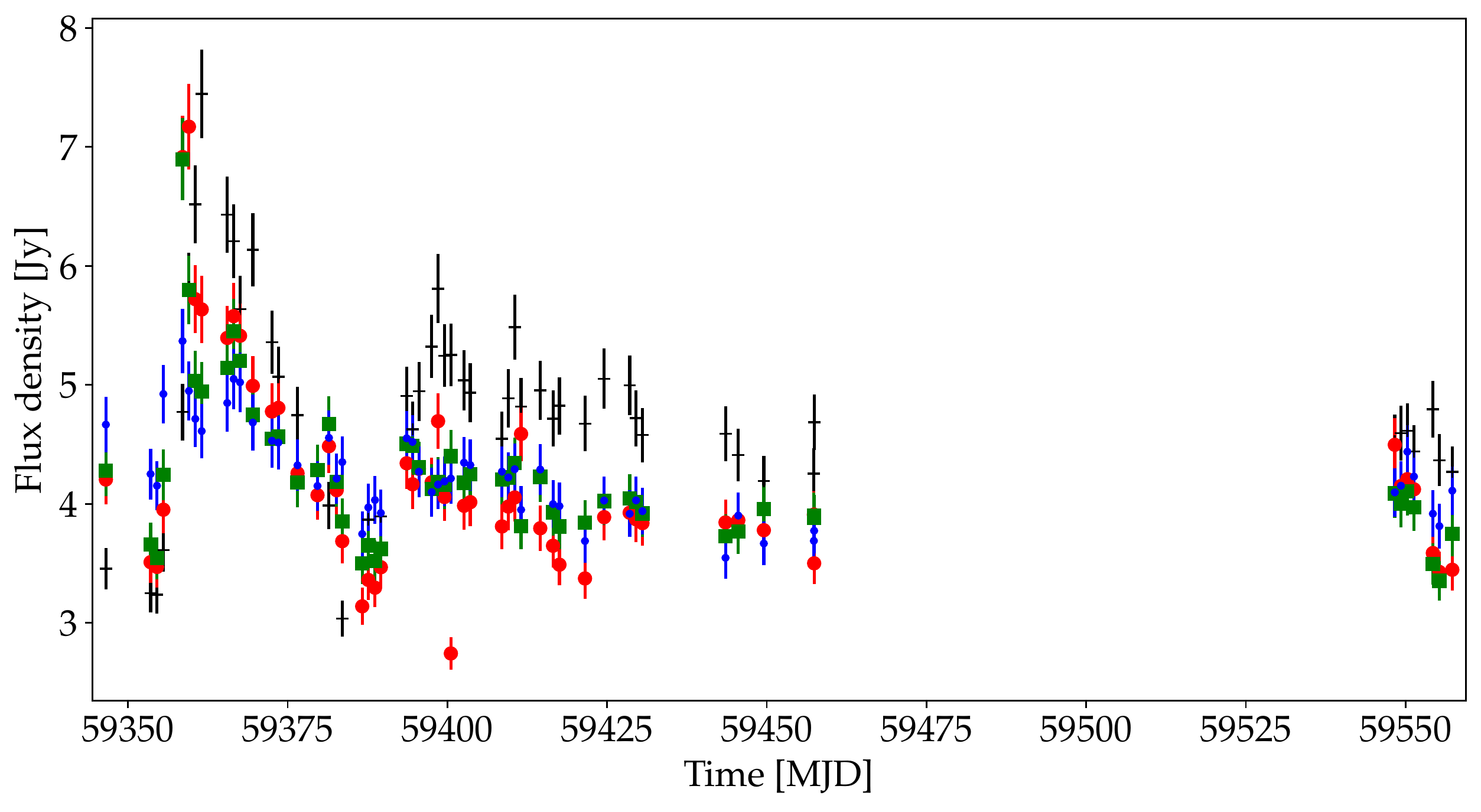}
     \caption{The multi-frequency light curve of the source 0059+581, as observed in the FM-sessions, shown separately to appreciate the rich variability. Flux densities in the four VGOS bands 1, 2, 3 and 4 are shown with black crosses, red circles, green squares and blue dots respectively. }
     \label{fig:fm_0059}
\end{figure*}

\begin{figure*}[htbp]
     \centering
     \begin{subfigure}[bt]{0.32\textwidth}
         \includegraphics[width=\textwidth]{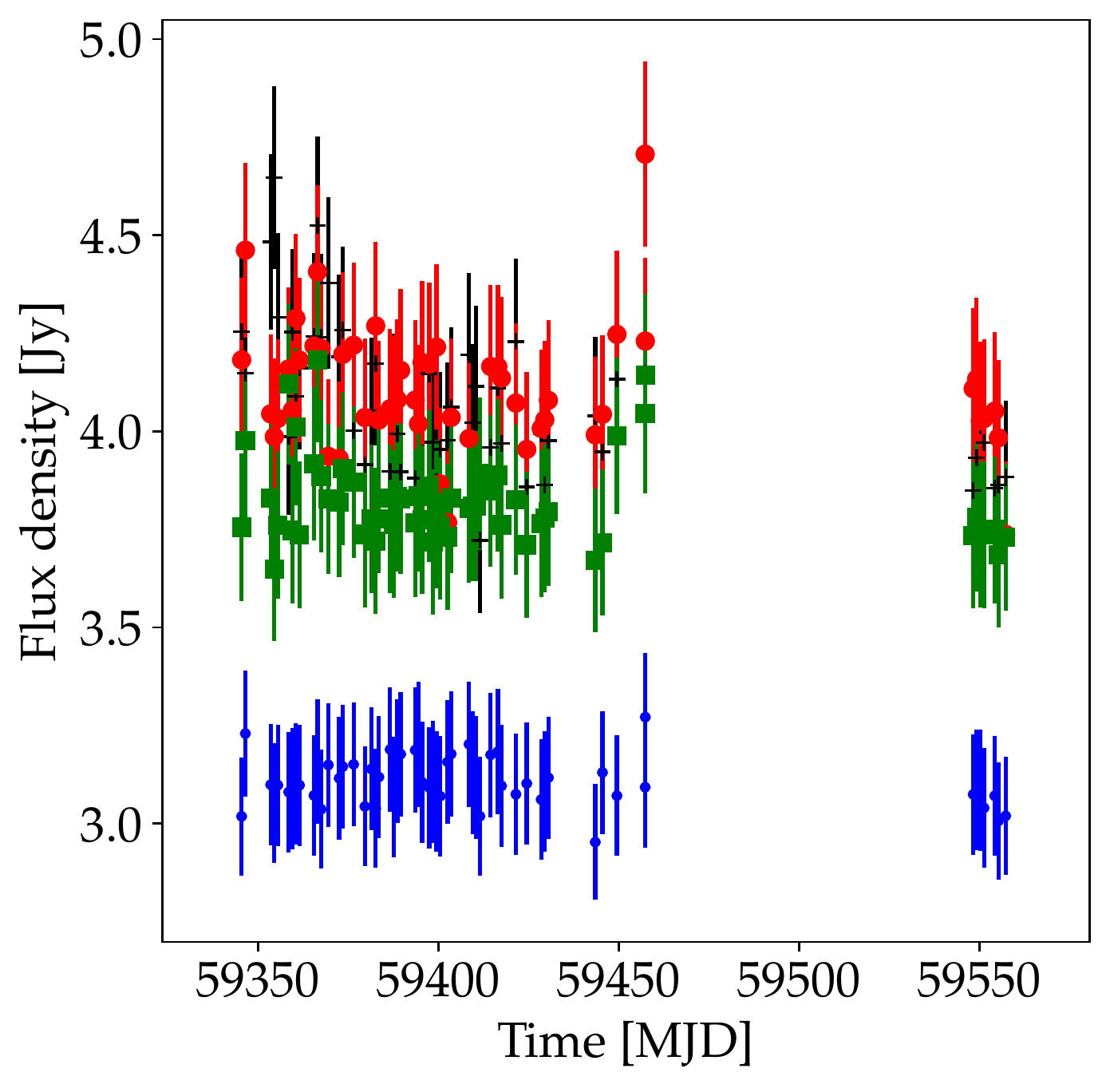}
         \caption{0552+398}
         \label{fig:fm_0552+398}
     \end{subfigure}
     \hfill
     \begin{subfigure}[bt]{0.32\textwidth}
         \includegraphics[width=\textwidth]{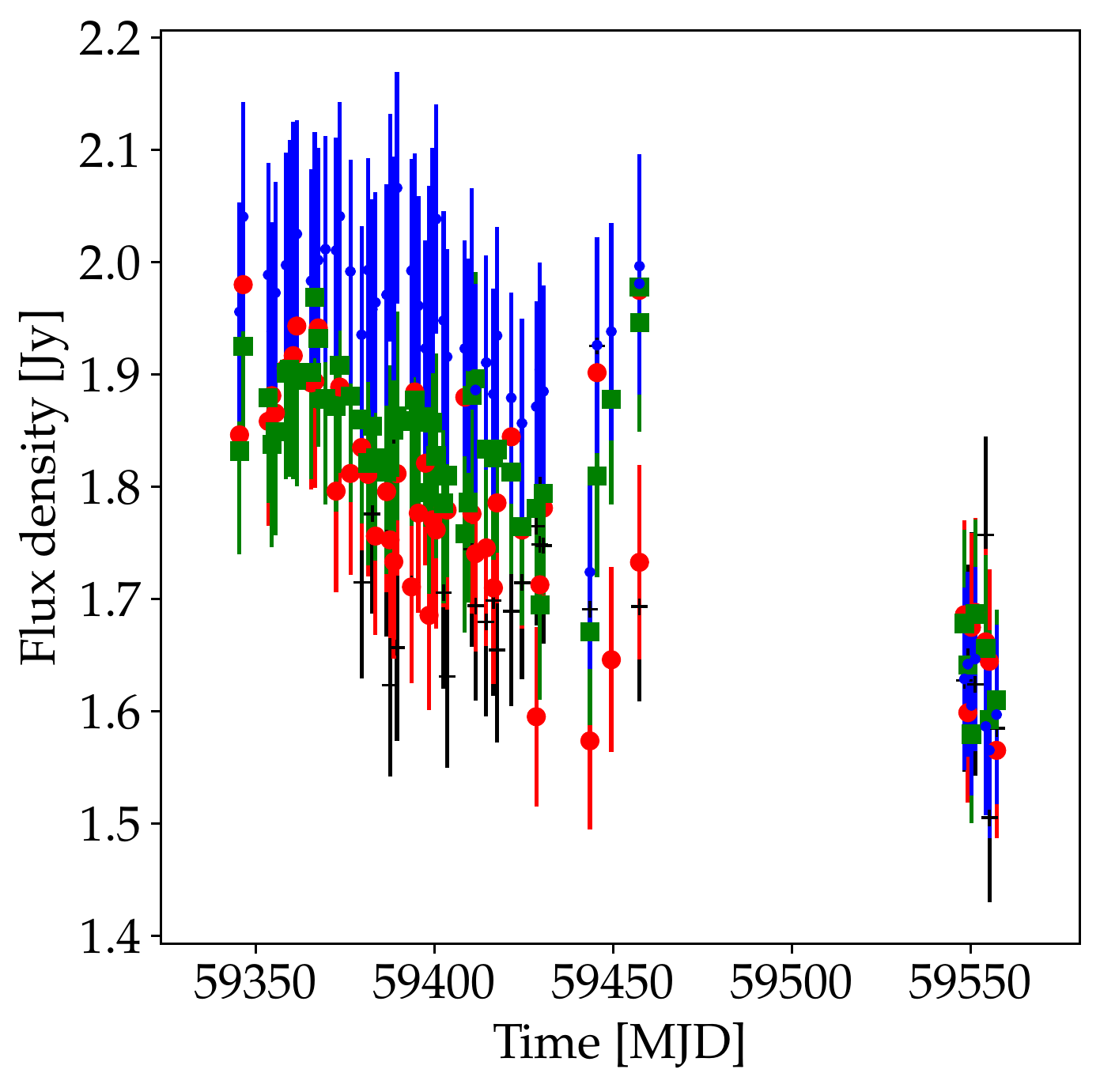}
         \caption{1144+402}
         \label{fig:fm_1144+402}
     \end{subfigure}
     \hfill
     \begin{subfigure}[bt]{0.32\textwidth}
         \includegraphics[width=\textwidth]{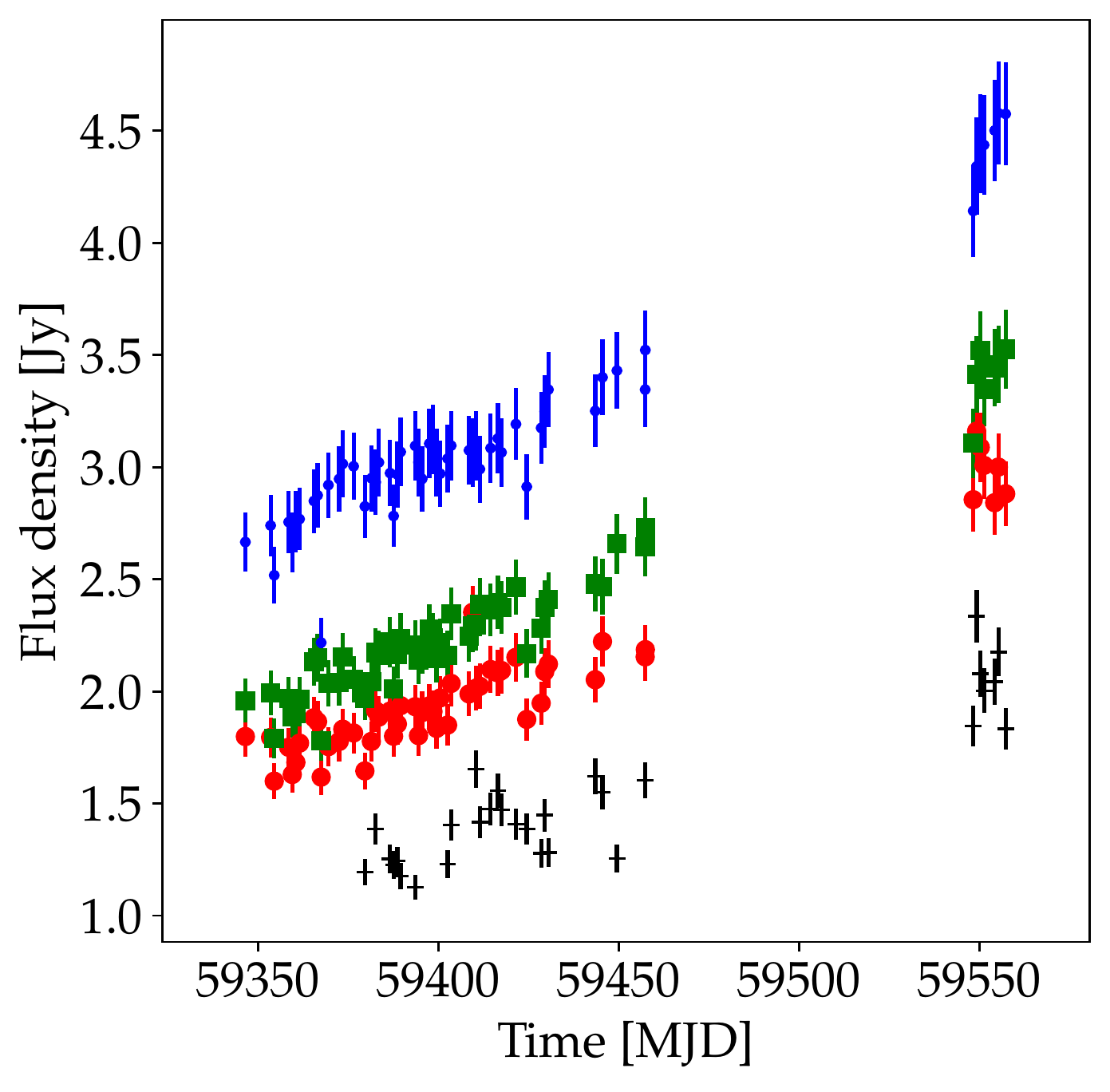}
         \caption{1156+295}
         \label{fig:fm_1156+295}
     \end{subfigure}
          \hfill
     \begin{subfigure}[bt]{0.32\textwidth}
         \includegraphics[width=\textwidth]{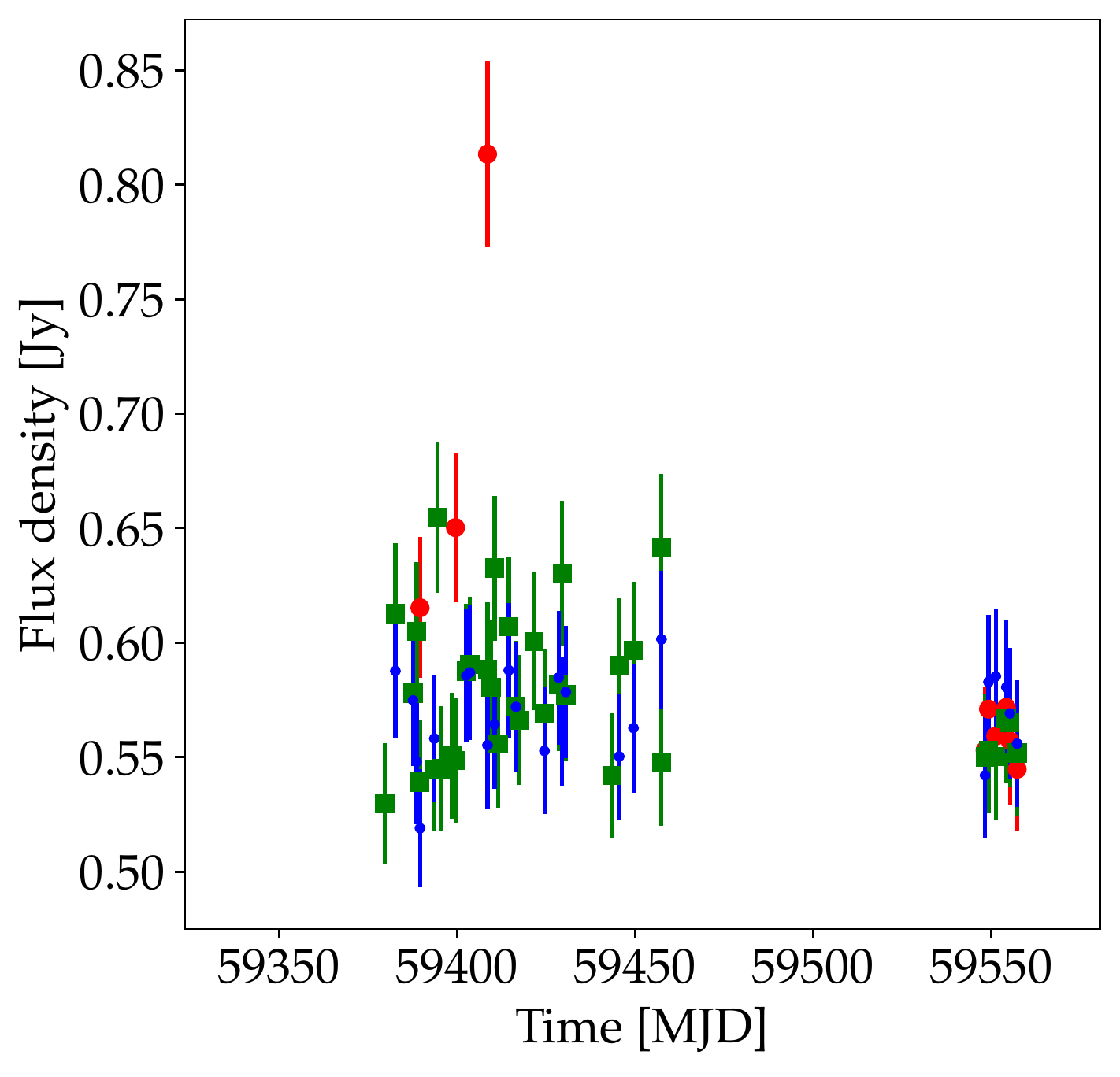}
         \caption{1617+229}
         \label{fig:fm_1617+229}
     \end{subfigure}
     \begin{subfigure}[bt]{0.32\textwidth}
         \includegraphics[width=\textwidth]{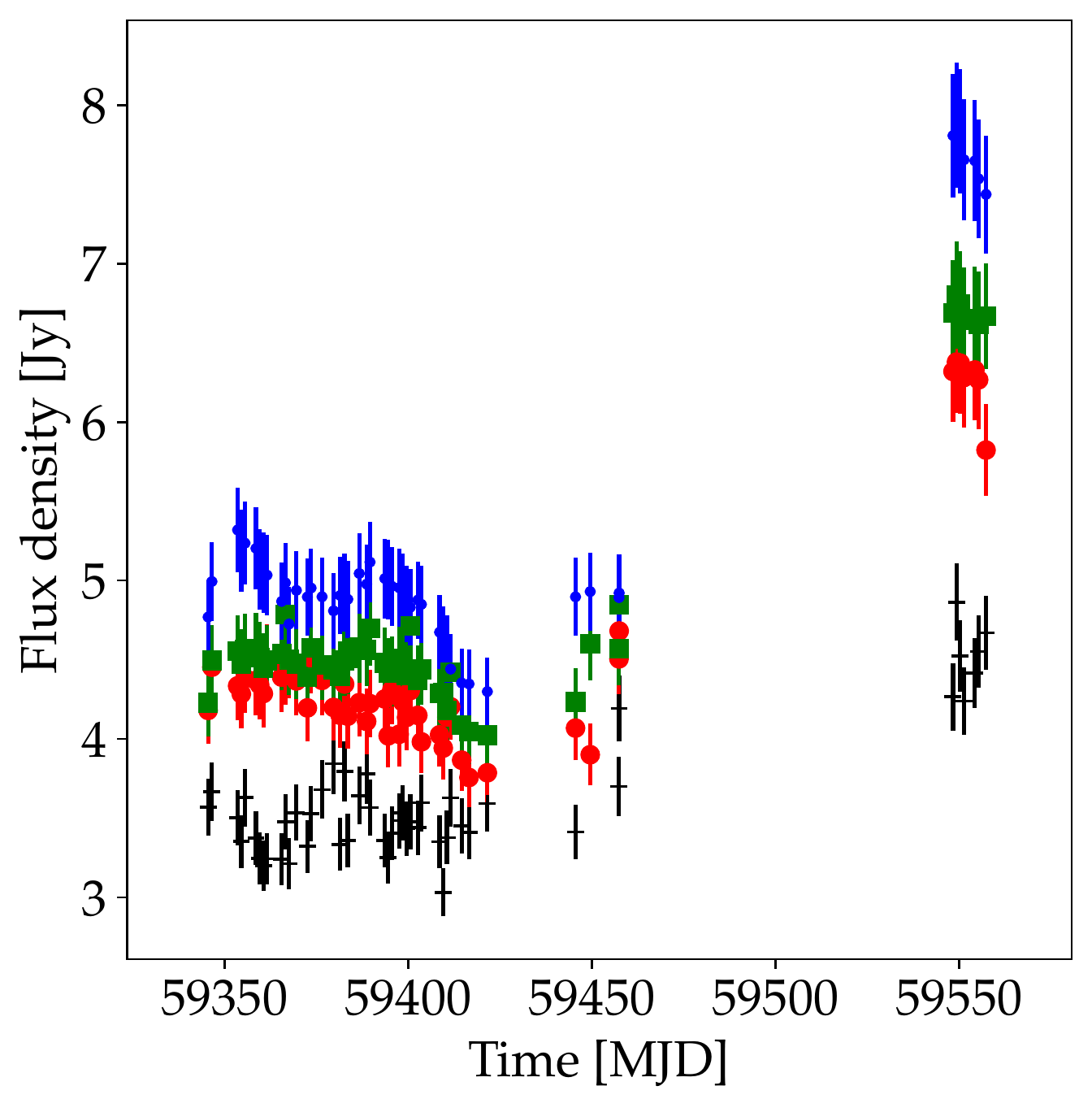}
         \caption{OJ287}
         \label{fig:fm_OJ287}
     \end{subfigure}
     \begin{subfigure}[bt]{0.32\textwidth}
         \includegraphics[width=\textwidth]{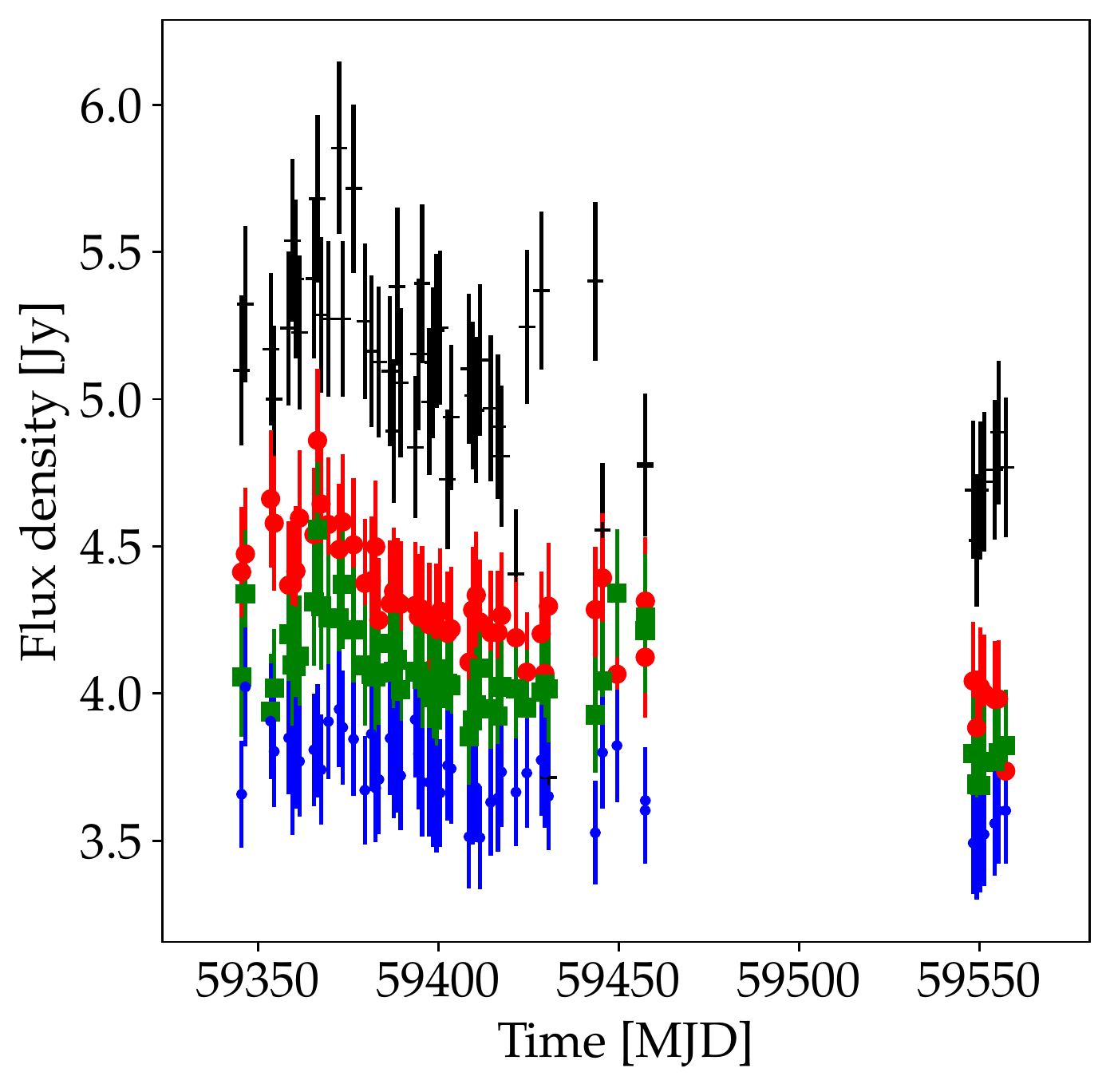}
         \caption{3C418}
         \label{fig:fm_3C418}
     \end{subfigure}
        \caption{Multi-frequency light curves of six sources observed in the FM-sessions. Flux densities in the four VGOS bands 1, 2, 3 and 4 are shown with black crosses, red circles, green squares and blue dots respectively. Note that the vertical scale is different for the different sources.}
        \label{fig:fm_quasars}
\end{figure*}

\begin{figure*}[htbp]
     \centering
        \begin{subfigure}[bt]{0.48\textwidth}
         \includegraphics[width=\textwidth]{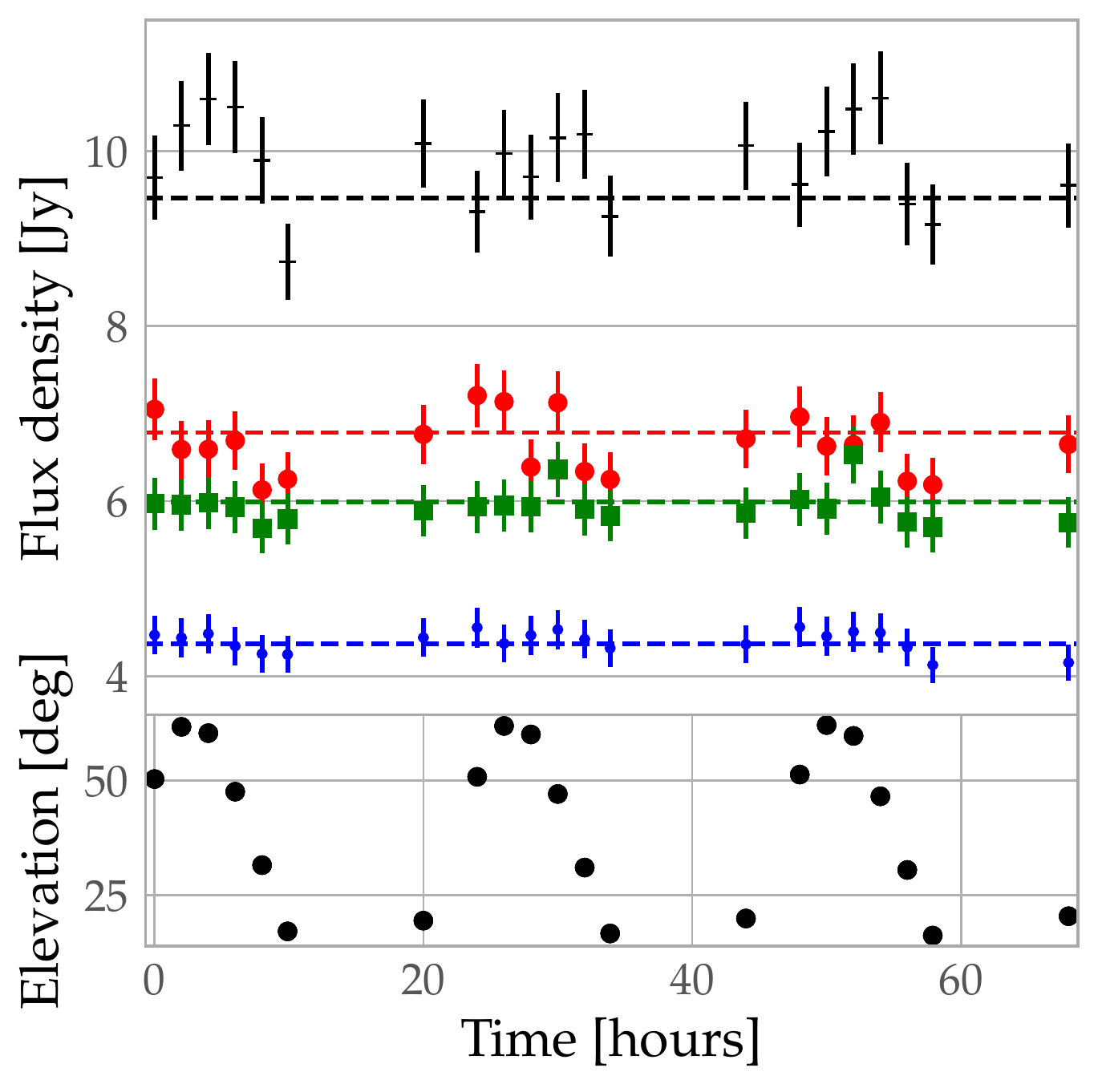}
         \caption{3C286}
         \label{fig:idvm_3C286}
     \end{subfigure}
          \hfill
     \begin{subfigure}[bt]{0.48\textwidth}
         \includegraphics[width=\textwidth]{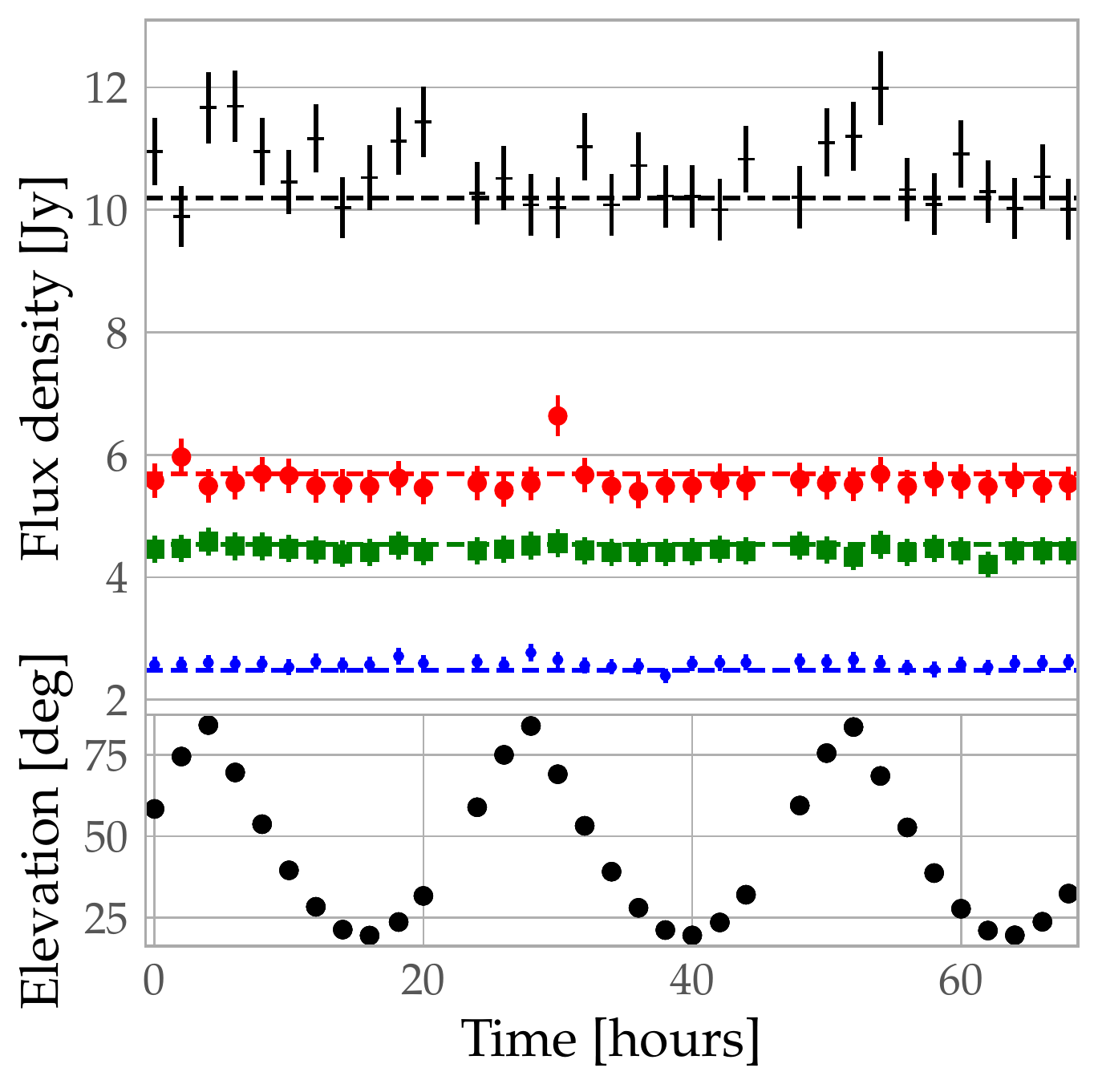}
         \caption{3C295}
         \label{fig:idvm_3C295}
     \end{subfigure}
     \begin{subfigure}[bt]{0.48\textwidth}
         \includegraphics[width=\textwidth]{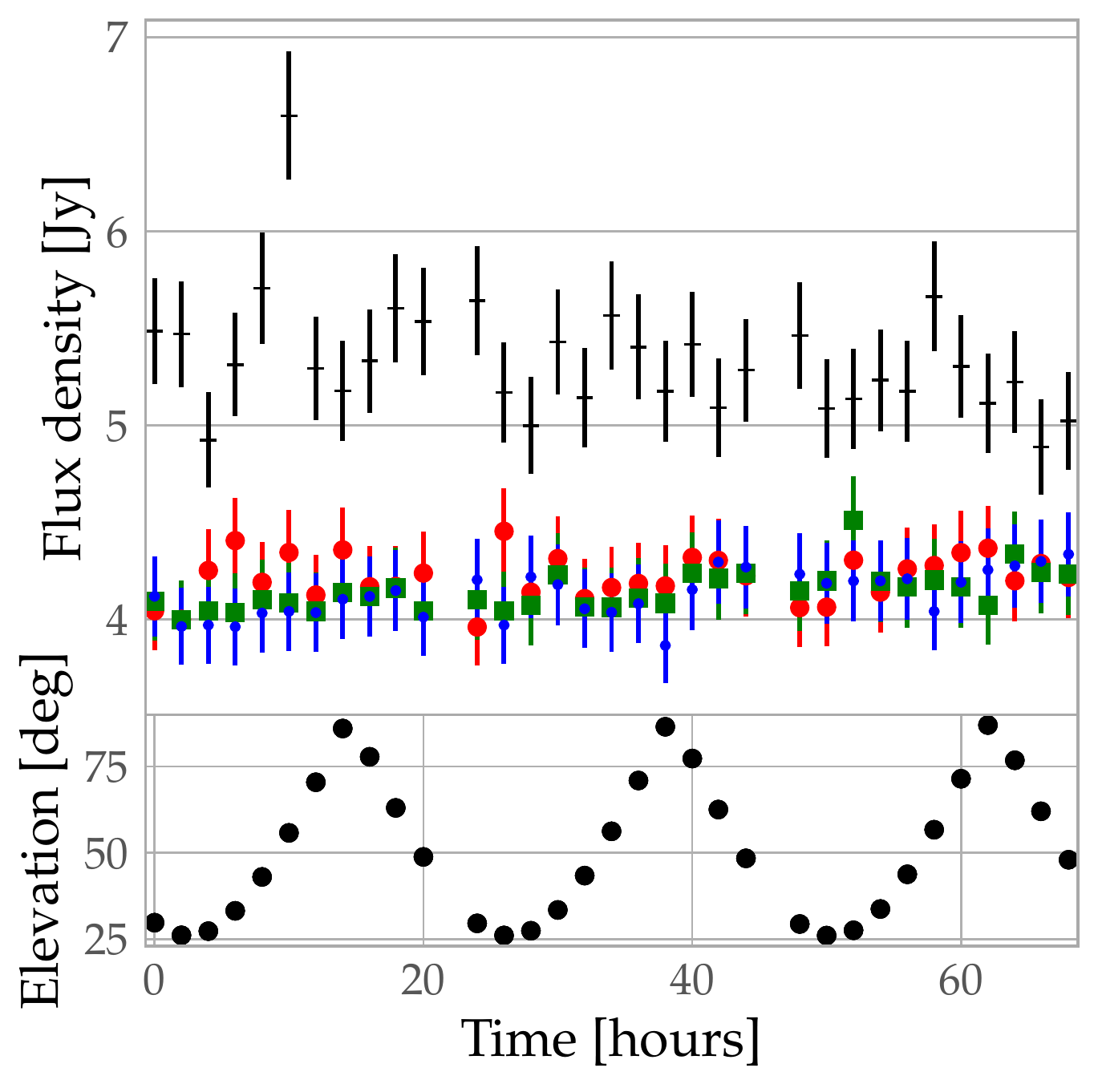}
         \caption{0059+581}
         \label{fig:idvm_0059+581}
     \end{subfigure}
     \hfill
     \begin{subfigure}[bt]{0.48\textwidth}
         \includegraphics[width=\textwidth]{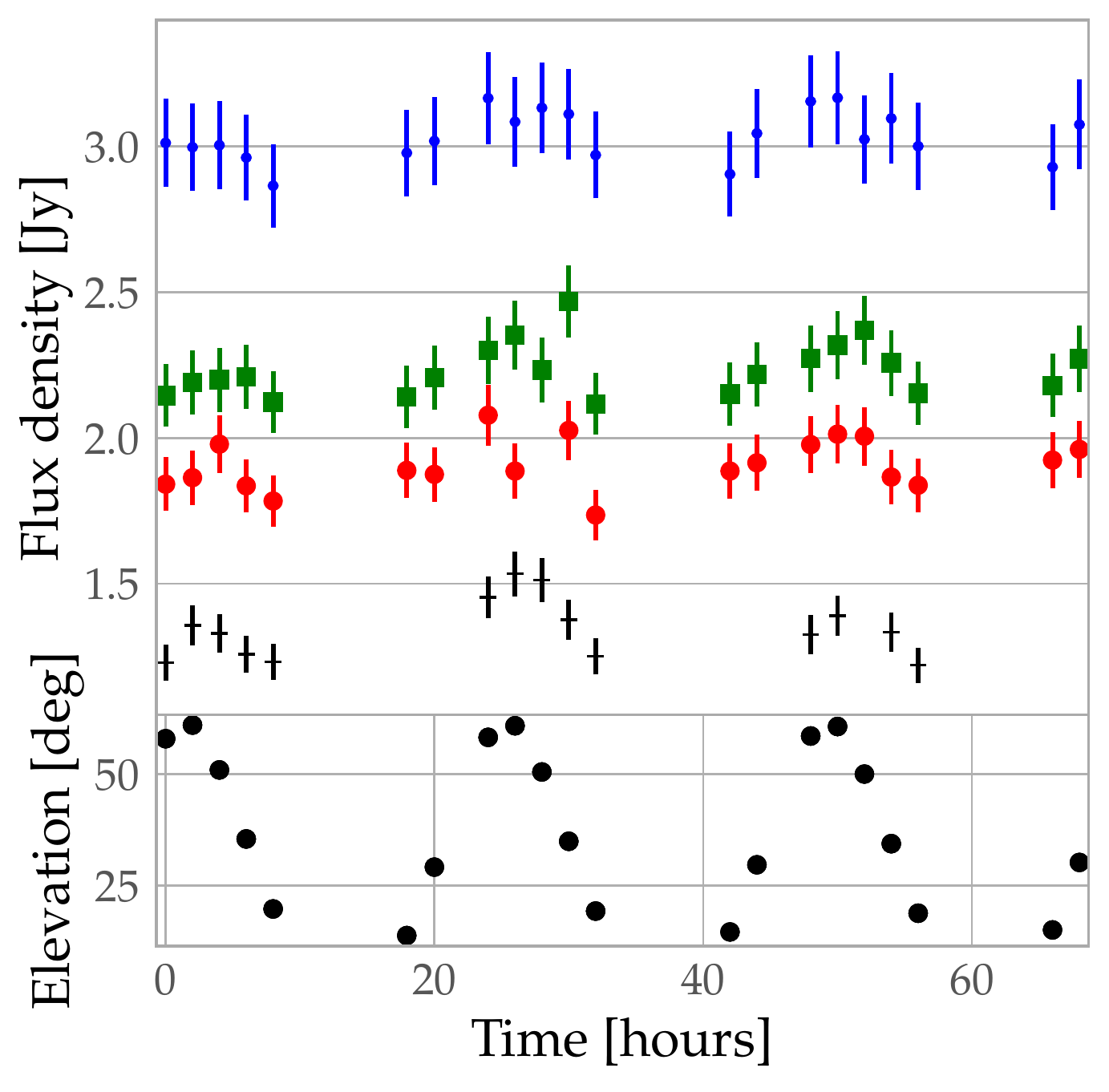}
         \caption{1156+295}
         \label{fig:idvm_1156+295}
     \end{subfigure}
     \hfill
        \caption{The top part of each panel show the multi-frequency light curves of the four sources observed in the IDVM-sessions. The bottom part of each panel shows the local elevation angle at observing time. Flux densities in the four VGOS bands 1, 2, 3 and 4 are shown with black crosses, red circles, green squares and blue dots respectively. Note that the vertical scale is different for the different sources. There is a slight correlation between the flux density and the elevation on the 5~\%-level. The starting time, given as 0 on the horizontal axis above, was MJD 59397.626.}
        \label{fig:idvm_fluxcal}
\end{figure*}

After processing the data using the methods and corrections described in Sect. \ref{sec:obs}, we obtained one Stokes I flux density value per source per experiment. These values are shown in Figs. \ref{fig:fm_fluxcal}, \ref{fig:fm_0059} and \ref{fig:fm_quasars} for the FM-experiments, and in Fig.~\ref{fig:idvm_fluxcal} for the IDVM experiments. Machine-readable tables with all values used to make the figures are available in the online material attached to this paper.

\section{Discussion}
\label{sec:discussion}
In this section we discuss the limitations of this study, and compare our results with other data from literature.
\subsection{Known limitations in scheduling and analysis}
In this work we were limited to short experiments, i.e. a limited number of sources, due to limited processing (correlation) power. We used a single 10-core machine, where each experiment took a few hours to process. This enabled one FM-session in the afternoon, and correlation in the evening or following day, before the next session. Some IDVM sessions, which were more closely spaced in time, only included 2 sources to decrease processing time. With more computing power available in the future, we expect to be able to process, and hence also observe, more sources.

We formed Stokes I as the average of the parallell linear polarisation products, i.e. (HH+VV)/2. This does not correct for neither parallactic angle differences, nor instrumental leakage. In principle, this could be accounted for by using e.g. PolConvert \citep{marti-vidal_et_al_2016, marti-vidal_et_al_2021}, but this is beyond the scope of this work. Instead we note that since the telescopes are so close, the differential parallactic angle is negligible. Furthermore, we empirically find that (for most SPWs and sources), the cross-polarisation products are negligible compared to HH and VV, implying that any instrumental leakage is not significant.

In this work we conservatively scheduled relatively long scans of 180~s for the flux density reference sources 3C286, 3C295, 3C147, 120~s for the weak 1617+229, and 60~s for the remaining six sources. We find, however, that fringes can be obtained in 30 sec solution intervals for $>$1 Jy level sources in 32 MHz bandwidth. This means we could significantly shorten the scan length if combining multiple spectral windows together in fringe-fitting. This is routinely done in geodetic VGOS processing, by correcting for instrumental phase- and delay offsets for each SPW so they can all be aligned for the broad-band fringe-fitting. In principle, both instrumental SPW-specific phase- and delay offsets, and directional-dependent offsets tracked using the pulse-calibration system, could be incorporated in our analysis. However, these SPW-specific corrections are challenging to apply in a in CASA, since there are not yet standard tasks to include e.g. pulse-calibration data. Since the single-SPW fitting worked sufficiently well to fulfil our aims, we did not pursue the issue of combining multiple SPWs further.

Finally, our RFI-mitigation using \emph{rflag} and filtering SPWs as outliers could be improved. Still, the strategy employed here appears to give robust results in all four VGOS bands.

\subsection{Notes on individual sources}
 We find our sources to be a mix of both stable and variable in time, as well as a range of spectral signatures. In this section we briefly discuss our results for the seven sources in relation to existing literature.

\subsubsection{0059+581}
0059+581 has been used as a bright and compact fringe finder during VGOS commissioning work. It does however exhibit significant flux density variability on short and long timescales, and has been included in multiple monitoring campaigns at radio \citep{Lister_et_al_2018, Lister_et_al_2021}, gamma-ray \citep{Paliya_et_al_2021} and optical \citep{Blinov_et_al_2021} wavelengths. This source was the one most often included in our monitoring, already from the start and our FM-results are shown in Fig.~\ref{fig:fm_0059}. We detect at least one apparent flare of emission, in all four bands, with the peak around MJD~59361. We do not detect any significant short-term variability in our IDMV sessions, presented in Fig.~\ref{fig:idvm_0059+581}. We find our data to be consistent with a flat spectrum, where the tentative excess emission at the lowest band~1 could be due to other nearby synchrotron sources in the field.

\subsubsection{OJ287}
OJ287 may host a supermassive black hole binary system \citep{Dey_et_al_2021}. It is known to be variable and has been monitored at multiple wavelengths for over a century (see e.g. \citealt{Hudec_et_al_2013, Komossa_et_al_2020, Lister_et_al_2021} and references therein). Our measurements are presented in Figs. \ref{fig:fm_OJ287}. We find OJ287 to exhibit some moderate variability during the first months. However, the significant increase in power observed in our final data points, taken in December 2021, suggest that it has entered an active phase. Continued monitoring of this source at VGOS frequencies could give valuable data for future multi-wavelengths studies.

\subsubsection{1156+295}
This source has also been included in monitoring campaigns, e.g. \cite{Lister_et_al_2019}, and studied with dedicated VLBI observations \citep{Zhao_et_al_2011}. As evident from Fig.~\ref{fig:fm_1156+295}, the flux density of 1156+295 increases by a factor 2 in a linear fashion during the 7 months spanned by our monitoring. The four bands follow each other, keeping the inverted (more power at higher frequencies) spectrum from start to end. We do not detect significant variability in our short IDVM sessions, see Fig.~\ref{fig:idvm_1156+295}.

\subsubsection{1617+229}
The blazar 1617+229 has also been included in monitoring campaigns \citep{Lister_et_al_2019,Richards_et_al_2014}. It is the weakest source in our sample, with a relatively stable and flat spectrum of about 0.5 Jy as can be seen in Fig.~\ref{fig:fm_1617+229}. We note that after the maintenance work (Sect. \ref{sec:gcomo}) it appears that we obtain robust estimates of the flux density also in band~2. Although our measurements are internally consistent, and consistent with literature, the results - in particular for the lowest band~1 - could be improved with refined calibration procedures (e.g. combining multiple SPWs in fringe-fitting).

\subsubsection{1144+402}
1144+402 is identified as a flat-spectrum radio source in the MOJAVE monitoring campaign \citep{Lister_et_al_2021}, in good agreement with our results presented in Fig.~\ref{fig:fm_1144+402}. We find this source to be stable over our measurement period, with similar flux densities obtained in all four bands.

\subsubsection{3C418}
3C418 has significant structure both on large (NVSS) and small (VLBI) scales and of interest to astronomers for many years (e.g. \citealt{Muxlow_et_al_1984}). It has been observed frequently in VGOS sessions, and it has been demonstrated that the structure significantly impacts geodetic data analysis \citep{Xu_et_al_2021}. We find this source to be relatively stable, see Fig.~\ref{fig:fm_3C418}, with band~2, 3, 4 flux densities of about 4~Jy. This is consistent with e.g. the S/X-band monitoring from the BVID, which shows a slow increase over time.

\subsubsection{0552+398}
This source has also been the subject of VLBI observations for many years (e.g. \citealt{Spangler_et_al_1983, Lister_et_al_2019}). We find it to be slowly varying with a flat spectrum in bands 1, 2 and 3. The multi-frequency lightcurve given in Fig.~\ref{fig:fm_0552+398} is consistent with the past monitoring data available via the BVID. 

\section{Conclusions}
\label{sec:conclusions}
In this work, we have arrived at the following conclusions:
   \begin{enumerate}
      \item The electronics and calibration tables used to monitor system temperatures for the Onsala twin telescopes are working as expected within the nominal 10~\% uncertainty.
      \item With minor improvements to the model of antenna gain vs. elevation, and corrections for backend systematics, flux densities of $\sim 500$~mJy-sources could in the future be monitored to  within a few \% in the 3 to 15~GHz band.
      \item We observed a bright multi-frequency flare in the source 0059+581.
      \item Once the full international VGOS network is capable of routinely monitoring system temperatures during observations, the astronomical community may get regular broad band flux-density monitoring (and mapping) of hundreds of sources "for free" during geodetic VLBI observations.
   \end{enumerate}

\backmatter

\bmhead{Supplementary information}
In addition to the manuscript, we will provide for electronic storage:
 \begin{enumerate}
  \item ASCII-text files with full-polarisation flux density measurements obtained for all analysed observations.
  \item ASCII-text files with Stokes I flux density values used to create the figures in the paper.
  \item A python script used to obtain the full-polarisation flux density measurements from the FITSIDI+antab data, as explained in the paper.
 \end{enumerate}

\bmhead{Acknowledgments}
The authors thank Leonid Petrov for constructive comments on a preliminary version of this manuscript.

\section*{Declarations}

\subsection*{Funding}
F. M. acknowledges funding from the Chalmers Astrophysics and Space Sciences Summer (CASSUM) research program.
\subsection*{Conflict of interest/Competing interests}
Not applicable.
\subsection*{Ethics approval}
Not applicable.
\subsection*{Consent to participate}
Not applicable.
\subsection*{Consent for publication} 
All authors approve the publication of this study.
\subsection*{Availability of data and material}
The full-polarisation (HH,HV,VH,VV) and Stokes I flux densities obtained and analysed in this study are included in the supplementary information files of this article.
The FITS and antab files generated and analysed in this study are available from the corresponding author on reasonable request.
\subsection*{Code availability}
The python-script used to analyse the FITS and antab data with CASA is included in the supplementary information files of this article.
\subsection*{Authors' contributions}
All authors contributed to the study conception and design. All authors contributed to the planning of observations. Eskil Varenius, Karine Le Bail, and R\"udiger Haas carried out the observations. Eskil Varenius and R\"udiger Haas correlated the raw voltage data. Eskil Varenius and Francesco Maio analysed the data using CASA. The first draft of the manuscript was written by Eskil Varenius and all authors commented on previous versions of the manuscript. All authors contributed to the interpretation of the data. All authors read and approved the final manuscript.

\bibliography{main}

\end{document}